\documentstyle[aps,pra,epsfig,twocolumn]{revtex}

\def\be{\begin{equation}}
\def\ee{\end{equation}}
\def\bea{\begin{eqnarray}}
\def\eea{\end{eqnarray}}
\def\bma{\begin{mathletters}}
\def\ema{\end{mathletters}}
\newcommand{\one}{\mbox{$1 \hspace{-1.0mm}  {\bf l}$}}
\newcommand{\eins}{\mbox{$1 \hspace{-1.0mm}  {\bf l}$}}
\def\C{\hbox{$\mit I$\kern-.7em$\mit C$}}

\newcommand{\ket}[1]{ | \, #1  \rangle}

\def\II{I(\{p_k\},\{\rho_k\})}
\def\ss{{\cal K}}

\begin{document}
\draft

\title{Visible compression of commuting mixed states}

\author{W. D\"ur, G. Vidal and J.I. Cirac}

\address
{Institut f\"ur Theoretische Physik, Universit\"at
Innsbruck,A-6020 Innsbruck, Austria}

\date{\today}

\maketitle

\begin{abstract} 
We analyze the problem of quantum data compression of commuting density 
operators in the visible case. We show that the lower bound for the compression 
factor given by the Levitin--Holevo function is reached by providing an 
explicit protocol.  
\end{abstract} 


\narrowtext

\section{Introduction}

The applications of Quantum Mechanics in the fields of communication, 
computation, and precision measurements are based on the possibility of encoding 
and manipulating information using quantum states. Thus, one of the most 
relevant questions in this context is the extension of Shannon's noiseless 
coding theorem \cite{Sh48} to the quantum domain. That is, to find out the 
minimum amount of resources needed for a faithful storage (encoding) and 
retrieval (decoding) of quantum states, or, equivalently, the most economical 
way of compressing them. For pure states, this problem was stated and solved by 
Schumacher \cite{Sc95,Jo94,Ba96}. For mixed states, however, this is still an 
open problem \cite{Ho98,Ba00,Ho00}.

The problem of quantum data compression can be formulated as
follows. Alice has a (stationary memoryless) quantum source that
produces systems in the state (described by the density
operator) $\rho_k$ with probability $p_k$, where $k=1,2,\ldots,
L$ ($L$ finite). Let us consider a sequence $\ss$ of $N$ systems
which, for simplicity, we consider to be qubits, created by the
source. Let us denote by $\sigma_{\ss}^A\equiv\rho_{k_1}\otimes
\rho_{k_2}\ldots \otimes \rho_{k_N}$ the corresponding
state \cite{note1}. Alice wants to transmit such a state to Bob
by using as few qubits as possible. That is: (i) she {\em
encodes} the sequence in a set of $M$ qubits (i.e., with the
help of her sequence she prepares them in some state) and sends
them to Bob; (ii) he {\em decodes} the state (i.e. with the help
of the qubits he has received he prepares a sequence of $N$
systems in some state $\sigma_{\ss}^B$). The goal is to find the
procedure for which, for sufficiently long sequences, Bob's
state $\sigma_{\ss}^B$ is ``arbitrarily close" to
$\sigma_{\ss}^A$ and, at the same time, $M$ is minimal
(arbitrarily close means with respect to some measure of
fidelity, see below). The quantity $C=\lim_{N\to\infty}M/N$ is
called compression factor.

In the case where the $\rho_k$ correspond to pure states one
finds that $C= S(\rho)$ \cite{Sc95}, where
\be
\rho\equiv\sum_{k=1}^L p_k \rho_k,
\ee
and
\be
S(\rho)\equiv -{\rm tr}[\rho \log_2(\rho)],
\ee
is the von Neumann entropy of $\rho$. When the $\rho_k$
correspond to mixed states, however, the value of $C$ is not
known (except for the somehow simple case in which the supports
of the operators $\rho_k$ are orthogonal \cite{Lo95}). It can be
shown that $S(\rho)\ge C \ge \II$ \cite{Ho98,Ba00}, where
\be\label{LH}
\II=S(\rho)-\sum_k p_k S(\rho_k),
\ee
is the Levitin--Holevo function. In Ref. \cite{Ba00}, the
authors analyze several cases where they are able to show that
$S(\rho) > C$ by providing explicit protocols. However, none of
those protocols achieve the lower bound $\II$. Thus, the
question whether this limit can be reached or not is still open.
In fact, it has been argued \cite{Ho00} that in the affirmative
case one could assign a definite meaning to the Levitin--Holevo
function besides the well known one related to the maximum
amount of classical information that can be stored and retrieved
in and from quantum states \cite{Ha96,Ho98b,Sc97}.

There are two different scenarios where quantum data compression
of mixed states has been analyzed \cite{Ho98,Ba00,Ho00}. In the
so--called {\it visible} scenario, Alice knows the state
$\sigma_{\ss}^A$ she wants to compress. In the {\it blind} one,
she does not know it. Obviously, the compression factor in the
visible scenario is smaller than or equal to that in the latter
one. In particular, for pure states both compression factors
coincide \cite{Sc95}.

In this paper we study the compression of quantum mixed states
in the visible scenario, and in the case in which the operators
$\rho_k$ commute with each other. We provide an explicit
protocol which reaches the lower bound for the compression
factor, which implies that
\be
\label{Cfactor}
C = \II.
\ee
The basic idea to achieve such compression factor is to let
Alice and Bob change the encoding/decoding procedure randomly
from sequence to sequence. For that, we will assume that Alice
and Bob possess the same random number generator (or,
equivalently, that they share a list of random numbers). We will
concentrate on the case in which the systems under consideration
are qubits. As we will indicate, the generalization to higher
dimensional systems is straightforward. Note that, as shown in
Ref. \cite{Ba00}, the problem analyzed in this paper is
equivalent to the one of classical data compression of
probability distributions. We will nevertheless use a quantum
mechanical language in view of a possible extension of our
protocol to the case in which the operators $\rho_k$ do not
commute. On the other hand, the reason why our protocol achieves
the compression factor (\ref{Cfactor}) can be easily understood
in terms of typical subspaces (or typical sequences in the
classical case). Thus, we will first explain how our protocol
works by using this concept. Once this is clear, a detailed
proof can be easily constructed.
It has come to our attention that \cite{Sa01} presents an alternative proof of 
the achievability of (\ref{Cfactor}) using rate distortion theory.

This paper is organized as follows. In Section II we qualitatively explain our 
protocol using the concept of typical sequences. In Section III we describe in 
detail our protocol for the case of two states ($L=2$) and show that it achieves 
the compression factor (\ref{Cfactor}). The protocol can be straightforwardly 
generalized to $L>2$ by following the ideas of Section II. However, we do not 
include the detailed proofs here since they require an involved notation, and do 
not add any new idea to the problem. In Section IV we discuss possible 
extensions of our protocol. Finally, the Appendix is concerned with some 
technical details.

\section{Description in terms of typical sequences}

In this section we formulate the problem in terms of typical
sequences, which allows us to explain the basic idea of our
protocol. We assume that Alice wants to send a sequence of $N$
qubits to Bob, each one in state $\rho_k$ with probability
$p_k$, where all the $\rho_k$ commute. We can always write
\be
\label{rhok}
\rho_k = \lambda_k |1\rangle\langle 1| +
(1-\lambda_k) |0\rangle\langle 0|,
\ee
Thus, we have
\be
\rho=\sum_{k=1}^L p_k \rho_k = \overline P_1 |1\rangle\langle 1| +
\overline P_0 |0\rangle\langle 0|,
\ee
where
\be
\overline P_1 =\sum_{k=1}^L p_k \lambda_k, \quad \overline P_0
=1-\overline P_1.
\ee
These quantities are the probability that the quantum source
creates the state $|1\rangle$ and $|0\rangle$, respectively.

As mentioned in the introduction, the goal is to compress a
sequence of the form $\sigma_{\ss}^A\equiv\rho_{k_1}\otimes
\rho_{k_2}\ldots \otimes \rho_{k_N}$, where $k_i=1,2,\ldots,L$.
We will denote by $v_k$ a vector whose elements indicate the
positions at which the operator $\rho_k$ appears. For example,
if we take the sequence
\be
\underbrace{\rho_1\otimes\rho_1\ldots\rho_1}_{n_1} \otimes
\underbrace{\rho_2\otimes\rho_2\ldots \rho_2}_{n_2} \otimes
\ldots \otimes \underbrace{\rho_L\otimes\rho_L\ldots \rho_L}_{n_L}
\ee
then $v_1=(1,2,\ldots,n_1)$, $v_2=(n_1+1,n_1+2,\ldots,n_1+n_2)$, etc.

If the sequence is sufficiently long, $\sigma_{\ss}^A$ will
contain the state $\rho_k$ approximately $\overline n_k\equiv
Np_k\gg 1$ times. Let us call a sequence which exactly contains
such a number of times these operators ``typical sequence''.
Moreover, since $\overline n_k\gg 1$ we can also apply the same
idea within the sequence that Alice wants to send. If we write
the operator $\sigma_{\ss}^A$ in the basis
$\{|i_1\rangle\otimes|i_2\rangle\ldots |i_N\rangle\}$
($i_j=0,1$), most of the contribution will come from states with
approximately $\overline n_k\lambda_k$ ones (and $\overline
n_k(1-\lambda_k)$ zeros) at the positions $v_k$. Let us call
``typical states'' those with exactly such numbers of zeros and
ones at the positions specified by $v_k$. Thus, let us
concentrate on a method in which, given a typical sequence,
Alice sends Bob enough information so that he can create at
random one of the corresponding typical states. It is
intuitively clear that if Alice can accomplish this task with
$M\sim N\II$ qubits, then she will also be able to send most of
the sequences with this amount of qubits and high fidelity.

So, let us now assume that Alice and Bob use their random number
generator to create {\em the same} random state of $N$ qubits,
each of them in the state $|0\rangle$ or $|1\rangle$ according
to the probabilities $\overline P_0$ and $\overline P_1$,
respectively. Let us denote by $p$ the probability that such a
state is a typical one for a given typical sequence. In that
case, if they create (instead of one) $\sim 1/p$ such random
states, the probability that among them there is a typical one
will be very close to one. In that case, Alice just has to tell
Bob which of those states randomly generated is the one that
corresponds to the typical sequence she is intending to send.
The number of qubits to give that information to Bob is
$M=\log_2(1/p)$. Since
\be
p = \label{prob}\frac{
\left(\begin{array}{c} \overline n_1 \\ \overline n_1\lambda_1 \end{array}\right)
\left(\begin{array}{c} \overline n_2 \\ \overline n_2\lambda_2 \end{array}\right)
\ldots
\left(\begin{array}{c} \overline n_L \\ \overline n_L\lambda_L \end{array}\right)}
{\left(\begin{array}{c} N \\ N\overline P_1 \end{array}\right)}
\ee
we obtain that $M=\log_2(1/p)\sim N\II$ (for $N\gg 1$).

\section{Protocol for two states}

In this Section we give the protocol to achieve the compression
factor (\ref{Cfactor}). We will show that for any $\epsilon,\delta>0$
there exists an $N_0$ such that the sequences with $N>N_0$
qubits can be encoded in $N[I(p_k,\rho_k)+\delta]$ qubits with
a fidelity $\overline F>1-\epsilon$. Here, $\overline F$ is the
averaged fidelity
\be
\overline F = \sum_{\ss} P_{\ss}
F(\sigma_{\ss}^A,\sigma_{\ss}^B),
\ee
$P_{\ss}$ is the probability that Alice sends the sequence
$\ss$, and \cite{note2}
\be
F(\sigma_1,\sigma_2) \equiv {\rm tr} \left[\sigma_1^{1/2}
\sigma_2
\sigma_1^{1/2}
\right]^{1/2}=
{\rm tr} [\sigma_1^{1/2}\sigma_2^{1/2}],
\ee
where the last equality holds for commuting operators.

We will concentrate in the case where there are only two
possible states ($L=2$). The general case can be analyzed in the
same way as here, although the notation becomes much more
involved. Thus, let us assume that Alice wants to send the
sequence $\ss$, consisting of $N$ qubits in states $\rho_1$ or
$\rho_2$, to Bob. As before, we will call $n_{1,2}$ (where
$n_2=N-n_1$) the number of times the operator $\rho_{1,2}$
appears in the sequence, and $v_{1,2}$ the positions where it
appears. Note that in all these quantities we should write a
subscript $\ss$ indicating their dependence on the particular
sequence Alice is trying to send. In order to keep the notation
simple, and whenever it is clear from the context, we will omit
in all the quantities the dependence on the particular sequence.
On the other hand, in the protocol given below we will consider
that Alice sends classical bits to Bob. Obviously, these
classical bits can in turn be encoded in the same number of
qubits if we choose the states $|0\rangle$ and $|1\rangle$. The
protocol consists of the following encoding and decoding
procedures:

\begin{enumerate}

\item Encoding:
\begin{enumerate}

\item Alice selects two integer numbers $x_1$ and $x_2$, with
$0\le x_{i} \le n_{i}$ according
to the following binomial distributions
\be
\label{Px12}
P(x_{i}) = \left(\begin{array}{c} n_{i} \\ x_{i}
\end{array}\right)
\lambda_{i}^{x_{i}} (1-\lambda_{i})^{n_{i}-x_{i}},
\ee
where $i=1,2$.

\item Using the common random number generator Alice creates
$S$ random sequences of $N$ bits each. Each of the bits is set
to 1 or 0 according to the
probability $P_1=(\lambda_1 n_1+\lambda_2 n_2)/N$, $P_0=1-P_1$,
respectively. She associates a number between $1$ and $S$ with
each sequence.

\item If among the $S$ sequences there are one or more
with exactly $x_{i}$ ones and $n_{i}-x_{i}$ zeros at the
positions indicated by $v_{i}$ for both $i=1,2$, then she
chooses one of them randomly and sends the number associated
with that sequence to Bob. Otherwise, she sends the number 0
(which indicates an error). Note that for that she uses
$[\log_2(S+1)+1]$ bits, where $[\ldots]$ denotes the integer
part.

\item She also encodes in a set of $[\log_2{N}+1]$ bits the value of
$n_{1}$ and sends it to Bob.
\end{enumerate}

\item Decoding

\begin{enumerate}

\item Bob uses the random number generator to create the same $S$ random
sequences as Alice and assigns the same numbers. Note that Bob
knows the values $n_{1}$ (since it has been sent by Alice)
and $n_2=N-n_1$.

\item Using the bits sent by Alice, he identifies the random sequence
and prepares $N$ qubits in the corresponding state (i.e.
prepares the qubits in states $|0\rangle$ or $|1\rangle$ if the
sequence contains a zero or a one at each position). If he
receives the error state, he prepares the qubits in a fixed
state $\sigma_0=\one/2^N$.
\end{enumerate}
\end{enumerate}

Before showing that the above protocol achieves the desired
bound, let us make some remarks. Firstly, we can replace the
condition imposed by $\delta>0$ on the number of bits needed to
encode the sequences by requiring that
\be
\label{cond1}
\log_2(S)=N[I(p_k,\rho_k)+f_N],
\ee
where $f_N\to 0$ as $N\to
\infty$. Note that the number of bits needed to transmit the
value of $n_1$ can be included in $f_N$ since $\log_2(N+1)/N\to
0$, and therefore need not be considered. Actually, one can
devise a similar encoding and decoding scheme in which this
number need not be transmitted. However, our scheme allows for a
simpler proof of our statements. Secondly, as it is shown in the
Appendix, we can replace the condition imposed by $\epsilon$ on
the averaged fidelity by
\be
\label{cond2}
E\equiv\sum_{n_1=0}^N P_{n_1} E_{n_1} <\epsilon
\ee
where $P_{n_1}$ is the probability that we have a sequence with
exactly $n_1$ times $\rho_1$ and the rest $\rho_2$, and
$E_{n_1}$ is the probability that Alice sends the error bit $0$
if she had one of such sequences. Thirdly, we will deal with
several binomial distributions, which have the form
\be
Q_y\equiv \left(\begin{array}{c} n \\ y
\end{array}\right)
p^{y} (1-p)^{n-y},
\ee
where $0<p<1$. We will use the following properties of such
distribution: (i) for all $\epsilon>0$ and $0<\eta<1/2$, there
exists some $n_0$ such that if $n>n_0$ then
\be
\sum_{y=[pn-n^{1/2+\eta}]}^{[pn+n^{1/2+\eta}]} Q_y> 1-\epsilon.
\ee
This property allows us to restrict the allowed values of the
parameters. For the sake of definiteness we will take
$\eta=0.1$. (ii) For $n$ sufficiently large and
$y\in[pn-n^{1/2+\eta},pn+n^{1/2+\eta}]$
\be
\label{proptwo}
Q_y > \frac{1}{2}
\frac{e^{-(y-np)^2/[2(np(1-p)]}}{\sqrt{2\pi np(1-p)}}.
\ee

Now, let us show that the protocol given above fulfills the
desired properties. First, given the fact that $P_{n_1}$ follows
a binomial distribution, we can restrict the summation in
(\ref{cond2}) to the values
\be
\label{n1}
n_1\in [\overline n_1- N^{1/2+0.1},\overline n_1+ N^{1/2+0.1}].
\ee
Moreover, the remaining sum is smaller than the maximum value of
$E_{n_1}$ where $n_1$ lies in the interval indicated in Eq.\
(\ref{n1}). This value can be determined with the help of Eq.\
(\ref{Ap5}). Since $P(x_{1,2})$, the probability that Alice
selects the values $x_1$ and $x_2$ in the step 1(a), is a
product of two binomial distributions, again for sufficiently
large $N$ we can restrict the sums to
\be
\label{x1}
x_{i}\in [n_{i}\lambda_{i}- N^{1/2+0.1}, n_{i}\lambda_{i}+
N^{1/2+0.1}], \quad i=1,2.
\ee

Thus, the problem is reduced to showing that for any
$\epsilon>0$, for sufficiently large $N$ we can choose $S$
fulfilling (\ref{cond1}) and so that the maximum value of
$E(x_{1,2},v_{1,2})$ with the restrictions (\ref{n1}) and
(\ref{x1}) is smaller than $\epsilon$, and where
$E(x_{1,2},v_{1,2})$ is the probability that the error state is
produced given the values of $x_{1,2}$ and $v_{1,2}$ (see
Appendix). We can always write
$E(x_{1,2},v_{1,2})=[1-R(x_{1,2},v_{1,2})]^S$, where
$R(x_{1,2},v_{1,2})$ is the probability that if we take a
sequence of zeros and ones according to the probabilities
$P_{1,0}$, the sequence exactly contains $x_{i}$ ones (and the
rest zeros) at positions $v_{i}$, for both $i=1,2$. Such a
probability can be calculated as
$R(x_{1,2},v_{1,2})=Q(x_1+x_2)P(x_{1,2},v_{1,2}/x_1+x_2)$, where
$Q(x_1+x_2)$ is the probability that the sequence contains
$x_1+x_2$ ones and $P(x_{1,2},v_{1,2}/x_1+x_2)$ is the
probability that those are at the correct positions. The first
one is given again by a binomial distribution; by using Eq.\
(\ref{proptwo}) one can easily find \cite{note4} that
\be
\label{Qbound}
Q(x_1+x_2)\ge K\frac{e^{-\alpha N^{0.2}}}{\sqrt{N}}
\equiv \frac{1}{a_N},
\ee
where $K$ and $\alpha$ are constants (independent of $N$). On
the other hand
\bea
P(x_{1,2},v_{1,2}/x_1+x_2) &=& \frac{
\left(\begin{array}{c} n_1 \\ n_1\lambda_1 \end{array}\right)
\left(\begin{array}{c} n_2 \\ n_2\lambda_2 \end{array}\right)}
{\left(\begin{array}{c} N \\ NP_1 \end{array}\right)}\nonumber\\
&\ge& 2^{-N\II-N^{1/2+0.2}}\equiv \frac{1}{b_N}
\eea
for sufficiently large $N$, as can be checked using the bounds
given by Stirling formulas. By choosing $S=Na_Nb_N$ we obtain
that $E(x_{1,2},v_{1,2})=[1-R(x_{1,2},v_{1,2})]^S\le
[1-1/(a_Nb_N)]^{Na_Nb_N}\to 0$ and (\ref{cond1}) with $f_N=
[\log_2(N)+o(N^{0.7})]/N\to 0$ for $N\to \infty$, as required.

\section{Possible extensions}

\subsection{$d$-level systems}

One can easily generalize our results to $d$--level systems. In that case, a 
quantum source produces $d$--level systems (qudits) in the state (described by 
the density operator) $\rho_k$ with probability $p_k$. For a faithful 
transmission of $N$ of those systems, $M$ qudits (equivalently 
$M\log_2(d)$ qubits) are required. In case all $\rho_k$ commute, the compression 
factor $C=\lim_{N\to\infty}M/N$ turns out to be $C=\II/\log_2(d)$, where $\II$ 
is given in (\ref{LH}) and the factor $\log_2(d)$ appears because we are dealing 
with $d$--level systems now. The number of qubits per signal states required for 
a faithful transmission is thus again given by the Levitin-Holevo function 
$\II$, so the lower bound can be reached also when dealing with $d$--level 
systems.

This can be understood qualitatevely in a similar way as in the qubit case (see 
Sec. II). The condition that all $\rho_k$ commute implies that we can always 
write 
\be
\rho_k=\sum_{j=1}^{d}\lambda_j^k|j\rangle\langle j|.
\ee
and thus
\be
\rho=\sum_{k=1}^{L} p_k \rho_k=\sum_{j=1}^{d} \overline{P_j}|j\rangle\langle j|,
\ee
where $\overline{P_j}=\sum_{k=1}^{L}p_k\lambda_j^k$. Proceeding in the same vain 
as in the qubit case, we find that the ``typical states'' of a certain (typical) 
sequence have exactly $Np_k\lambda_j^k$ states $|j\rangle$ at the positions 
$v_k$. It is straightforward to calculate the probability $p$ that a state of 
$N$ qudits generated randomly according to the probability distribution 
$\{\overline{P_i}\}$ is a typical one for a given sequence. One finds that $p$ 
is given by an expression which is similar to (\ref{prob}), however the binomial 
factors are replaced by multinomial factors. This is due to the fact that the 
corresponding distributions are now multinomial instead of binomial. The number 
of required qubits, $M\log_2(d)$, turns out to be $\log_2(1/p)\sim N\II$ (for 
$N\gg 1$), which leads to the announced compression factor. Also the detailed 
proof can be carried out in a similar way, replacing the binomial distributions 
by multionomial distributions and the corresponding Gaussian curves (see e.g. 
(\ref{proptwo})) by multidimensional Gaussians curves.

\subsection{Decoding without knowing the source}

Notice that in our protocol for compressing commuting mixed states we have 
implicitly assumed, in step 2(b) of the decoding stage, that Bob knows which are 
the eigenvectors of the density matrices, i.e. $\ket{0}$ and $\ket{1}$. This is 
of course legitimate in any context where both Alice and Bob are provided with a 
description of the source.

Let us note here that we can slightly modify the protocol in such a way that it 
works even if Bob does not have such a description. Indeed, suppose that now the 
eigenstates are $\ket{0'}$ and $\ket{1'}$. All we need is that Alice uses the 
quantum channel to sent $N$ copies of each of these states. Since the $N$ copies 
of (say) $\ket{0'}$, $\ket{0'}^{\otimes N}$, are supported on the 
(N+1-dimensional) symmetric subspace of $N$ qubits, $[log (N+1)]$ qubits are 
sufficient to transmit them. For large $N$, this does not change the 
communication cost per qubit, $\II$. And thus, once Bob has received and 
decompressed $\ket{0'}^{\otimes N}$ and $\ket{1'}^{\otimes N}$, he can use 
single copies of these states to replace the $\ket{0}$ and $\ket{1}$'s of step 
2(b). In this way, he does not need to know the details of the source to prepare 
faithful sequences $\sigma_{\ss}^B$.

\subsection{The Levitin-Holevo bound can not always be reached in a blind protocol}

In the previous sections, we showed for commuting density operators that in the 
visible scenario, the bound for the compression factor given by the 
Levitin--Holevo function can always be reached. Here, we investigate the {\it 
invisible} scenario, i.e. the case where Alice does not know the specific 
sequence to be sent. We give an example where the Levitin-Holevo bound for the 
compression factor cannot be reached. 

We consider two density operators $\rho_1=|1\rangle\langle1|$, $\rho_2=1/2\eins$ 
with corresponding probabilities $p_1=p_2=1/2$. We will argue that the 
achievable compression factor $C$ is given by the entropy of the operator 
$\rho=\sum_kp_k\rho_k$, $S(\rho) \approx 0.8113$, which should be compared with 
$\II \approx 0.3113$. We will not give a formal proof of this statement, but 
will rather argue in terms of typical sequences and the corresponding ``typical 
states'' (see Sec. II).

If we write the operator $\sigma_{\ss}^A$ corresponding to a typical sequence in 
the basis $\{|i_1\rangle\otimes|i_2\rangle\ldots |i_N\rangle\}$ ($i_j=0,1$), the 
typical states are those with exactly $3N/4$ ones (and $N/4$ zeros). Note that 
Alice can determine with help of a measurement of all qubits in the 
computational basis which of the typical states she possesses. This can be done 
without disturbing the signal because she measures in the eigenbasis of 
$\sigma_{\ss}^A$. Let us thus assume that Alice knows the typical state she has 
to transmit. We can take without loss of generality the state
\be
|a\rangle=
\underbrace{|1\rangle\otimes|1\rangle\otimes\ldots\otimes|1\rangle\otimes|1\rangle}_{3N/4} \otimes
\underbrace{|0\rangle\otimes\ldots\otimes|0\rangle}_{N/4}
\ee 
However, in contrast to the visible case, Alice does not know to which specific 
(typical) sequence the state $|a\rangle$ belongs to. In fact, there are many 
sequences which are compatible with the state $|a\rangle$, namely all those 
which have all $N/2$ density operators $\rho_1$ at the positions 
$1,\ldots,3N/4$. 

We will show now that the state $|a\rangle$ has to be transmitted ``perfectly'' 
to Bob, since even a small derivation from the state $|a\rangle$  will lead to a 
macroscopic error. To this aim, we consider a general coding/decoding procedure. 
Notice that Bob can measure in the computational basis after decoding the 
received signal and thereby obtain with some probability a pure state 
$|b\rangle$ \cite{foot6} which is a sequence of zeros and ones. Let us assume 
that $|b\rangle$ differs from $|a\rangle$ only at two positions, e.g. the first 
and the $N^{\rm th}$ qubits are flipped (note that two states must always differ 
at an even number of positions, as we assumed that the total number of 
zeros/ones is fixed). The average error can be written as follows
\be
E=\sum P(\sigma_{\ss}^A/a) E(b,\sigma_{\ss}^A),
\ee
where the sum runs over all possible typical sequences $\sigma_{\ss}^A$, 
$P(\sigma_{\ss}^A/a)$ is the probability that we deal with sequence 
$\sigma_{\ss}^A$ provided that Alice possesses the state $|a\rangle$ and 
$E(b,\sigma_{\ss}^A)$ is the error for the sequence $\sigma_{\ss}^A$ given that 
Bob received the state $|b\rangle$. Under our previous assumption on $|a\rangle, 
|b\rangle$, we have that $E(b,\sigma_{\ss}^A)$ is either one (for all sequences 
which have $\rho_1$ at position one) or zero (for all sequence which have 
$\rho_2$ a position one). As there are 
\be
q\equiv \left(\begin{array}{c} 3N/4 \\ N/2 \end{array}\right)
\ee
sequences which are compatible with $|a\rangle$, we have that 
$P(\sigma_{\ss}^A/a)=1/q$ for all those sequences and zero otherwise. It is easy 
to see that 
\be
E=\left(\begin{array}{c} 3N/4-1 \\ N/2-1 \end{array}\right)/\left(\begin{array}{c} 3N/4 \\ N/2 \end{array}\right)=2/3,
\ee
i.e. the average error is already macroscopic even when $|b\rangle$ differs from 
$|a\rangle$ only at two positions. We conclude that in order to have $E$ 
sufficiently small (and thus the fidelity sufficiently close to one), we must 
have that $|b\rangle=|a\rangle$. This implies that {\it all} typical states have 
to be transmitted perfectly from Alice to Bob, as our analysis is not restricted 
to the specific choice of $|a\rangle$. There are
\be
g\equiv\left(\begin{array}{c} N \\ 3N/4 \end{array}\right)
\ee 
typical states, which means that $\log_2(g) \sim N S(\rho) \approx 0.8113 N$ 
qubits are required for perfect transmission and no further compression is 
possible \cite{note5}. Thus, the Levitin-Holevo bound cannot be reached in this 
case. On the other hand, if $p_1=\epsilon \rightarrow 0$, it happens that ---also in 
the invisible scenario--- the achievable compression factor approaches $\II 
\rightarrow 0$, while $S(\rho) \rightarrow 1$.

Note that this analysis is not restricted to this specific example but can be 
generalized to determine the compression factor $C$, $S(\rho) \geq C \geq \II$, 
also in the invisible case. 

\section{Summary}

We have analyzed the compression of mixed states in the visible
case and for commuting density operators. We have given a
protocol that achieves the compression factor (\ref{Cfactor}),
which was known to be a lower bound. Our protocol is based on
the creation of the same set of random numbers by Alice and Bob,
and choosing among them the one appropriated to the sequence
they want to send. This protocol can be extended to the case in
which the density operators do not commute. In that case, Alice
and Bob can encode the states in the same random subspaces
within the typical subspace. This problem will be addressed in a
future work.

\section*{Acknowledgments}
We thank C. Fuchs for interesting discussions.
This work was supported by the Austrian Science Foundation under
the SFB ``control and measurement of coherent quantum systems´´
(Project 11), the European Community under the TMR network
ERB--FMRX--CT96--0087 and project EQUIP (contract
IST-1999-11053), the European Science Foundation, and the
Institute for Quantum Information GmbH. G.V also
acknowledges funding from the EC through grant No. HPMF-CT-1999-00200.


\appendix

\section{Bob's density operator and Fidelity}

We denote by $\{|\Psi_m\rangle\}_{m=1}^{2^N}$ the computational
basis for the $N$ qubits of the sequence, i.e.
$|\Psi_1\rangle=|0,0,\ldots,0\rangle$,
\ldots,$|\Psi_{2^N}\rangle=|1,1,\ldots,1\rangle$.
According to the protocol given in Section III, Bob's density
operator can be written as follows:
\bea
\label{Ap1}
\sigma_{\ss}^B &=& \sum_{x_{1,2}=0}^{n_{1,2}}
P(x_{1,2}) \sum_{t=1}^S P(t,x_{1,2},v_{1,2})
\nonumber\\
&& \times\sum_{m=1}^{2^N} P(m/t,x_{1,2},v_{1,2})
|\Psi_m\rangle\langle\Psi_m|
\nonumber\\ &&+
\sum_{x_{1,2}=0}^{n_{1,2}} P(x_{1,2})
E(x_{1,2},v_{1,2})\frac{\one}{2^N}.
\eea
Here, $P(x_{1,2})$ is the probability that Alice obtains $x_{1}$
and $x_{2}$ and is given in (\ref{Px12}); $P(t,x_{1,2},v_{1,2})$
is the probability that among the $S$ random sequences, there
are $t$ with exactly $x_{1,2}$ ones (and the rest zeros) at the
positions indicated by $v_{1,2}$; $E(x_{1,2},v_{1,2})\equiv
P(0,x_{1,2},v_{1,2})$, i.e. the probability that the error state
is produced; $P(m/t,x_{1,2},v_{1,2})$ is the probability that
given $t$ sequences with exactly $x_{1,2}$ ones (and the rest
zeros) at the positions indicated by $v_{1,2}$, and we choose
one of them randomly, Bob obtains the sequence of zeros and ones
corresponding to $|\Psi_m\rangle$. This last can be reexpressed
as
\bea
P(m/t,x_{1,2},v_{1,2}) &=& \sum_{x=0}^t \left(\begin{array}{c}
t\\ x \end{array}\right) P(m/x_{1,2},v_{1,2})^x
\nonumber\\&&\times [1-P(m/x_{1,2},v_{1,2})]^{t-x}
\frac{x}{t}\nonumber\\ &=& P(m/x_{1,2},v_{1,2}),
\eea
where $P(m/x_{1,2},v_{1,2})\equiv P(m/1,x_{1,2},v_{1,2})$. Now,
we can perform the sum over $t$ in (\ref{Ap1}) and obtain
\bea
\label{Ap2}
\sigma_{\ss}^B &=& \sum_{x_{1,2}=0}^{n_{1,2}}
P(x_{1,2}) [1-E(x_{1,2},v_{1,2})]
\nonumber\\ &&\times\sum_{m=1}^{2^N} P(m/x_{1,2},v_{1,2})
|\Psi_m\rangle\langle\Psi_m|
\nonumber\\ &&+
\sum_{x_{1,2}=0}^{n_{1,2}} P(x_{1,2}) E(x_{1,2},v_{1,2})
\frac{\one}{2^N}.
\eea

On the other hand, we can write
\be
\sigma_{\ss}^A = \sum_{x_{1,2}=0}^{n_{1,2}}
P(x_{1,2}) \sum_{m=1}^{2^N} P(m/x_{1,2},v_{1,2})
|\Psi_m\rangle\langle\Psi_m|.
\ee

The fidelity $F(\sigma_{\ss}^A,\sigma_{\ss}^B)$ will be larger
than or equal to the one calculated by ignoring the term
proportional to the identity operator in (\ref{Ap2}). We obtain
\bea
F(\sigma_{\ss}^A,\sigma_{\ss}^B) &\ge&
\sum_{x_{1,2}=0}^{n_{1,2}}\sum_{m=1}^{2^N}
P(x_{1,2}) P(m/x_{1,2},v_{1,2})\nonumber\\ &&\times
[1-E(x_{1,2},v_{1,2})]^{1/2}
\nonumber\\
&\ge& 1 - E_{\ss},
\eea
where
\be
\label{Ap5}
E_{\ss}=\sum_{x_{1,2}=0}^{n_{1,2}} P(x_{1,2})
E(x_{1,2},v_{1,2}),
\ee
and we have used
\be
\sum_{m=1}^{2^N} P(m/x_{1,2},v_{1,2})=1.
\ee

Thus, the condition
\be
E=\sum_{\ss} P_{\ss} E_{\ss} < \epsilon,
\ee
automatically implies that $\overline F>1-\epsilon$. Now, both
$P_{\ss}$ and $E_{\ss}$ only depend on the number of times that
$\rho_1$ appears in $\ss$, and not on how they are placed, so
that we can write (\ref{cond2}).


\begin{references}
\bibitem{note1}
In the following we will freely interchange the words sequence
and state whenever there is a one to one correspondence.
\bibitem{note2}
Note that, according to Ref. \cite{Ba00} we are using the global
fidelity as opposed to the local fidelity. However, in the case
of commuting operators the result does not depend on the
fidelity (both for the blind and visible cases). This can be
easily understood as follows. If there exist a protocol which
achieves a smaller compression factor using the local fidelity,
then Alice and Bob can use the same protocol but in which Bob
measures the decoded sequence of qubits in the
$\{|0\rangle,|1\rangle\}$ basis. It is clear that this method
will not change the local fidelity, but will give a global
fidelity equal to the local one. On the other hand, as pointed
out in Ref.\ \cite{Ba00} the local fidelity criterion is less
stringent than the global one.
\bibitem{note3}
The values $n_{1,2}$ and $v_{1,2}$ depend on the sequence $\ss$.
However, in order to make the notation simpler we will omit this
dependence throughout.
\bibitem{note4}
Note that the specific values of $K$ and $\alpha$ are irrelevant
for our discussion. Nevertheless, using the extreme values of
$n_i$ and $x_i$ given in the intervals (\ref{n1}) and
(\ref{x1}), and bounding the resulting expressions, one can
easily find, for example, that (\ref{Qbound}) holds with
$K=(8\pi \overline P_1\overline P_0)^{1/2}$ and
$\alpha=72/(\overline P_1 \overline P_0)$.
\bibitem{foot6}

It follows from the increasing character of the fidelity under trace preserving 
operations \cite{Ni00}, that if we measure non--selectively the decoded state 
$\sigma_{\ss}^B$ in the computational basis, its fidelity with respect to the 
initial state $\sigma_{\ss}^A$ will not decrease (notice that the initial state 
$\sigma_{\ss}^A$ is left invariant under the measurement). If we now in addition 
make the same measurement with post selection, we obtain a certain pure state 
$|b\rangle$, which can be written as a sequence of zeros and ones.
\bibitem{note5}
Note that when using a different definition of the fidelity, further compression 
may be possible. Consider for example the average local fidelity, which is defined 
as $\overline F= 1/N\sum_{l=1}^N F(\rho_l,\hat\rho_l)$, where $\rho_l$ 
[$\hat\rho_l$] is the reduced density operator at position $l$ of the original 
[received] system (sequence). This definition of fidelity is different from 
both, global and local fidelity as used in \cite{Ba00} and throughout this 
paper. In that case, the error $E(b,\sigma_{\ss}^A)$ scales like $2/N$, as 
errors on two positions only affect two out of $N$ density operators. This 
suggests that one may allow for certain imperfection in the transmission of state 
$|a\rangle$ without producing a macroscopic error. Thus not all typical states 
have to be sent perfectly but one may rather achieve further compression.
\end{references}
\end{document}